\documentclass[twocolumn,prd,preprintnumbers,nofootinbib]{revtex4}

\usepackage{graphicx}
\usepackage{bm}
\usepackage{amssymb,amsmath}

\usepackage{array}

\begin{document}


\preprint{CALT-68-2603}
\preprint{KRL-MAP-321}

\title{Model Independent Bounds on Magnetic Moments of Majorana Neutrinos}
\author{Nicole F. Bell}
\author{Mikhail Gorchtein} 
\author{Michael J. Ramsey-Musolf}
\author{Petr Vogel}
\author{Peng Wang}
\affiliation{
California Institute of Technology, 
Pasadena, CA 91125, USA} 

\date{September 23, 2006}

\begin{abstract}
 

We analyze the implications of neutrino masses for the magnitude of
neutrino magnetic moments.  By considering electroweak radiative
corrections to the neutrino mass, we derive model-independent
naturalness upper bounds on neutrino magnetic moments, $\mu_\nu$,
generated by physics above the electroweak scale. For Dirac neutrinos,
the bound is several orders of magnitude more stringent than present
experimental limits. However, for Majorana neutrinos the magnetic
moment contribution to the mass is Yukawa suppressed. The bounds we
derive for magnetic moments of Majorana neutrinos are weaker than
present experimental limits if $\mu_\nu$ is generated by new physics
at $\sim$ 1 TeV, and surpass current experimental sensitivity only for
new physics scales $>$ 10 -- 100 TeV. The discovery of a neutrino
magnetic moment near present limits would thus signify that neutrinos
are Majorana particles.

\end{abstract}

\pacs{Valid PACS appear here}

\maketitle


\section{Introduction}

In the Standard Model (minimally extended to include non-zero neutrino
mass) the neutrino magnetic moment is given by~\cite{Marciano:1977wx}
\begin{equation}
\mu_\nu\approx 3\times 10^{-19}\left(\frac{m_\nu}{1{\rm eV}}\right)\mu_B. 
\label{SM}
\end{equation}
An experimental observation of a magnetic moment larger than that
given in Eq.(\ref{SM}) would be an unequivocal indication of physics
beyond the minimally extended Standard Model.  Current laboratory
limits are determined via neutrino-electron scattering at low
energies, with $\mu_\nu < 1.5 \times 10^{-10} \mu_B$~\cite{Beacom} and
$\mu_\nu < 0.7 \times 10^{-10} \mu_B$~\cite{reactor} obtained from
solar and reactor experiments, respectively.  Slightly stronger bounds
are obtained from astrophysics.  Constraints on energy loss from
astrophysical objects via the decay of plasmons into
$\nu\overline{\nu}$ pairs restricts the neutrino magnetic moment to be
$\mu_\nu < 3 \times 10^{-12}$~\cite{Raffelt}.  Neutrino magnetic
moments are reviewed in~\cite{Fukugita,boris,Wong}, and recent work
can be found in~\cite{McLaughlin,Balantekin:2006sw}.

\begin{figure}[th]
\includegraphics[width = 3in]{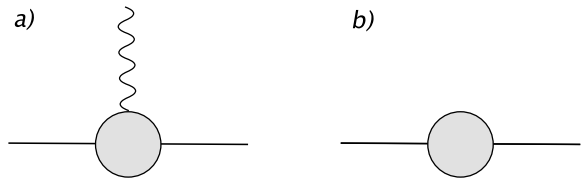}
\caption{a) Generic contribution to the neutrino magnetic moment
induced by physics beyond the standard model. b) Corresponding
contribution to the neutrino mass.  The solid and wavy lines
correspond to neutrinos and photons respectively, while the shaded
circle denotes physics beyond the SM.}
\label{fig:naive}
\end{figure}

It is possible to write down a na\"ive relationship between the size of
$\mu_\nu$ and $m_\nu$.  If a magnetic moment is generated by physics
beyond the Standard Model (SM) at an energy scale $\Lambda$, as in
Fig.~\ref{fig:naive}a, we can generically express its value as
\begin{equation}
\mu_\nu \sim \frac{eG}{\Lambda},
\end{equation}
where $e$ is the electric charge and $G$ contains a combination of
coupling constants and loop factors.  Removing the photon from the
same diagram (Fig.~\ref{fig:naive}b) gives a contribution to the
neutrino mass of order
\begin{equation}
m_\nu \sim G \Lambda.
\end{equation}
We thus have the relationship
\begin{eqnarray}
m_\nu &\sim&  \frac{\Lambda^2}{2 m_e}  \frac{\mu_\nu}{\mu_B} 
\nonumber \\
&\sim& \frac{\mu_\nu}{ 10^{-18} \mu_B}
[\Lambda({\rm TeV})]^2  \,\,\,\,{\rm eV},
\label{naive}
\end{eqnarray}
which implies that it is difficult to simultaneously reconcile a small
neutrino mass and a large magnetic moment.

However, it is well known that the na\"ive restriction given in
Eq.(\ref{naive}) can be overcome via a careful choice for the new
physics.  For example, we may impose a symmetry to enforce $m_\nu=0$
while allowing a non-zero value for
$\mu_\nu$~\cite{Voloshin,Georgi,Grimus,mohapatra1}, or employ a spin
suppression mechanism to keep $m_\nu$ small~\cite{Barr}.  Note
however, that these symmetries are typically broken by Standard Model
interactions.  The original version of the well-known Voloshin
mechanism~\cite{Voloshin} involved imposing an $SU(2)_\nu$ symmetry,
under which the left-handed neutrino and antineutrino ($\nu$ and
$\nu^c$) transform as a doublet.  The Dirac mass term transforms as a
triplet under this symmetry and is thus forbidden, while the magnetic
moment term is allowed as it transforms as a singlet.  However, the
$SU(2)_\nu$ symmetry is violated by SM gauge interactions.
For Majorana neutrinos, the Voloshin mechanism may be implemented
using flavor symmetries, such as those
in~\cite{Grimus,Georgi,mohapatra1}.  These flavor symmetries are not
broken by SM gauge interactions but are instead violated by SM Yukawa
interactions, provided that the charged lepton masses are generated
via the standard mechanism through Yukawa couplings to the SM Higgs
boson.\footnote{If the charged lepton masses are generated via a
non-standard mechanism, SM Yukawa interactions do not necessarily
violate flavor symmetries.  However, such flavor symmetries must always
be broken via some mechanism in order to obtain non-degenerate masses
for the charged leptons.}  By calculating neutrino magnetic moment
contributions to $m_\nu$ generated by SM radiative corrections, we may
thus obtain general, \lq\lq naturalness" upper limits on the size of
neutrino magnetic moments.

In the case of Dirac neutrinos, a magnetic moment term will
generically induce a radiative correction to the neutrino mass of
order~\cite{dirac}
\begin{eqnarray}
m_\nu &\sim& \frac{\alpha}{16\pi} 
\frac{\Lambda^2}{m_e}  \frac{\mu_\nu}{\mu_B} 
\nonumber \\
&\sim& \frac{\mu_\nu}{3 \times 10^{-15} \mu_B}
[\Lambda({\rm TeV})]^2 \,\,\,\,{\rm eV}.
\end{eqnarray}
If we take $\Lambda \simeq$ 1 TeV and $m_\nu \alt$ 0.3 eV, we obtain 
the limit $\mu_\nu \alt 10^{-15} \mu_B$, which is several orders of
magnitude more stringent than current experimental constraints.

The case of Majorana neutrinos is more subtle, due to the relative
flavor symmetries of $m_\nu$ and $\mu_\nu$ respectively.  The
transition magnetic moment $\left[\mu_\nu\right]_{\alpha\beta}$ is
antisymmetric in the flavor indices $\{\alpha,\beta\}$, while the mass
terms $[m_\nu]_{\alpha\beta}$ are symmetric.  These different flavor
symmetries play an important role in our limits, and are the origin of
the difference between the magnetic moment constraints for Dirac and Majorana
neutrinos~\cite{dirac}.

It has been shown in~\cite{Davidson} that the constraints on Majorana
neutrinos are significantly weaker than those obtained in~\cite{dirac}
for Dirac neutrinos, as the different flavor symmetries of $m_\nu$ and
$\mu_\nu$ lead to a mass term which is suppressed by charged lepton
masses.  This conclusion was reached by considering one-loop mixing of the
magnetic moment and mass operators generated by Standard Model interactions.
The authors of Ref.~\cite{Davidson} found that
if a magnetic moment arises through a coupling of the
neutrinos to the neutral component of the $SU(2)_L$ gauge boson, the
constraints obtained in~\cite{Davidson} for $\mu_{\tau e}$ and
$\mu_{\tau\mu}$ are comparable to present experiment limits, while the
constraint on $\mu_{e\mu}$ is significantly weaker.  Furthermore,
constraints on magnetic moments generated only via a coupling 
to the $U(1)_Y$ (hypercharge) gauge boson were found to be
additionally suppressed.  Thus, the analysis of~\cite{Davidson} lead to a model independent bound that is less stringent than
present experimental limits.

We shall show that two-loop matching of mass and magnetic moment operators implies stronger constraints than those obtained
in~\cite{Davidson} if the scale of the new physics $\Lambda \agt 10$
TeV. Moreover, these constraints apply to a magnetic moment generated
by either the hypercharge or $SU(2)_L$ gauge boson.
In arriving at these conclusions, we construct the most general set of operators that
contribute at lowest order to the mass and magnetic moments of
Majorana neutrinos, and derive model independent constraints which
link the two.   We thus obtain a
completely model independent naturalness bound that -- for $\Lambda \agt 100$
TeV -- is stronger than present experimental limits (even for the
weakest constrained element $\mu_{e\mu}$.)  Our key findings are
summarized in Table~\ref{summary}.

Our result implies that an experimental discovery of a magnetic moment
near the present limits would signify (i) neutrinos are Majorana
fermions and (ii) new lepton number violating physics responsible for
the generation of $\mu_\nu$ arises at a scale $\Lambda$ which is well
below the see-saw scale.  Implications for neutrinoless double beta
decay in theories with low scale lepton number violation have been
discussed in~\cite{Cirigliano}.


\section{Framework}

Following Refs.~\cite{dirac,Davidson} we assume that the magnetic
moment is generated by physics beyond the SM at an energy scale
$\Lambda$ above the electroweak scale.\footnote{We do not consider
models with new light particles, such as~\cite{mohapatra2}.}  In order
to be completely model independent, the new physics will be left
unspecified and we shall work exclusively with dimension $D>4$
operators involving only SM fields, obtained by integrating out the
physics above the scale $\Lambda$.  We thus consider an effective
theory that is valid below the scale $\Lambda$, respects the
$SU(2)_L\times U(1)_Y$ symmetry of the SM, and contains only SM fields
charged under these gauge groups.  
We assume that all the usual SM interactions are present, including the 
Yukawa couplings of the charged leptons to the SM Higgs.  
We shall also work with the flavor states $\nu_\alpha$ (i.e. the basis
where the charged lepton masses are diagonal) and discuss the mass
eigenstate basis in Section~\ref{flavor}. Note that in either basis,
Majorana neutrinos cannot have diagonal magnetic moments, but are
permitted non-zero transition moments.

The lowest order contribution to the neutrino (Majorana) mass arises
from the usual five dimensional operator containing Higgs and
left-handed lepton doublet fields, $L$ and $H$, respectively:
\begin{equation}
\left[O_M^{5D}\right]_{\alpha\beta}\,=\,
\left(\overline{L^c_\alpha}\epsilon H\right)\left(H^T\epsilon L_\beta\right),
\label{OM5}
\end{equation}
where $\epsilon = - i \tau_2$, $\overline{L^c}=L^TC$, $C$ denotes
charge conjugation, and $\alpha$, $\beta$ are flavor indices.  The neutrino magnetic moment operator must be
generated by gauge invariant operators involving the $SU(2)_L$ and
$U(1)_Y$ gauge fields, $W_\mu^a$ and $B_\mu$ respectively.  The lowest
order contribution arises at dimension seven, so we consider the
following operators,
\begin{eqnarray}
\left[O_B\right]_{\alpha\beta}&=& g' 
\left(\overline{L^c}_\alpha\epsilon H\right)\sigma^{\mu\nu}
\left(H^T\epsilon L_\beta\right)B_{\mu\nu}, 
\label{OB}\\
\left[O_W\right]_{\alpha\beta}&=& g
\left(\overline{L^c_\alpha}\epsilon H\right)\sigma^{\mu\nu}
\left(H^T\epsilon \tau^a L_\beta\right)W_{\mu\nu}^a,
\label{OW}
\end{eqnarray}
where $B_{\mu\nu} = \partial_\mu B_\nu - \partial_\nu B_\mu$ and
$W_{\mu\nu}^a = \partial_\mu W_\nu^a - \partial_\nu W_\mu^a - g
\epsilon_{abc}W_\mu^b W_\nu^c$ are the U(1)$_Y$ and SU(2)$_L$ field
strength tensors, respectively, and $g'$ and $g$ are the corresponding
couplings.
We also define a 7D mass operator as
\begin{equation}
\left[O_M^{7D}\right]_{\alpha\beta} = 
\left(\overline{L^c_\alpha}\epsilon H\right)
\left(H^T\epsilon L_\beta\right) \left(H^\dagger H \right).
\label{OM7}
\end{equation}
We note that the three 7D operators given by Eqs.(\ref{OB}-\ref{OM7}) do not form a closed set under renormalization.  The
full basis of 7D operators may be found in the appendix.

Operators $O_M^{5D}$ and $O_M^{7D}$ are flavor symmetric, while $O_B$ is
antisymmetric. The operator $O_W$ is the most general 7D operator
involving $W_{\mu\nu}^a$ (see the appendix for details).  However, as
it is neither flavor symmetric nor antisymmetric it is useful to
express it in terms of operators with explicit flavor symmetry,
$O_W^\pm$, which we define as
\begin{eqnarray}
\left[ O_W^\pm \right]_{\alpha\beta} &=& \frac{1}{2} \left\{
\left[O_W\right]_{\alpha\beta} \pm \left[O_W\right]_{\beta\alpha} \right\}.
\end{eqnarray}
Our effective Lagrangian is therefore
\begin{eqnarray}
{\cal L} &=& \frac{C_M^{5D}}{\Lambda} O_M^{5D} 
+ \frac{C_M^{7D}}{\Lambda^3} O_M^{7D} \\
\nonumber
&&+  \frac{C_{B}}{\Lambda^3} O_{B} 
+\frac{ C_{W}^+}{\Lambda^3} O_{W}^+ +  \frac{C_{W}^-}{\Lambda^3} O_{W}^-+\cdots \ \ ,
\end{eqnarray}
where the \lq\lq $+\cdots$" denote other terms that are not relevant to the present analysis.

Due to the differences in the flavor structure of $O_W^+$ and $O_W^-$, the phenomenological
manifestations of $C_W^+$ and $C_W^-$ are quite distinct.  After
spontaneous symmetry breaking, the flavor antisymmetric operators
$O_B$ and $O_W^-$ contribute to the magnetic moment interaction
\begin{equation} 
\frac{1}{2} \left[\mu_\nu\right]_{\alpha\beta}\, 
\overline{\nu^c}_\alpha \sigma^{\mu\nu}
\nu_\beta F_{\mu\nu},
\end{equation}
where $F_{\mu\nu}$ is the electromagnetic field strength tensor,  
\begin{equation}
\frac{\left[\mu_\nu\right]_{\alpha\beta}}{\mu_B} = \frac{2m_e v^2}{\Lambda^3} 
\left(\left[C_B(M_W)\right]_{\alpha\beta} 
+ \left[C_W^-(M_W)\right]_{\alpha\beta}\right),
\label{munu}
\end{equation}
and the Higgs vacuum expectation value is $\langle H^T \rangle =(0, v/\sqrt{2})$. The flavor symmetric operator
$O_W^+$ does not contribute to this interaction at tree-level.
Similarly, the operators $O_M^{5D}$ and $O_M^{7D}$ generate
contributions to the Majorana neutrino mass terms,
$\frac{1}{2}\left[m_\nu\right]_{\alpha\beta}\overline{\nu^c}_\alpha \nu_\beta$,
given by
\begin{equation}
\frac{1}{2}\left[ m_\nu \right]_{\alpha\beta} 
= \frac{v^2}{2 \Lambda} \left[C_M^{5D}(M_W)\right] 
+ \frac{v^4}{4 \Lambda^3} \left[C_M^{7D}(M_W)\right].
\label{mnu}
\end{equation}

In sections~\ref{su2} and \ref{u1b} below, we calculate radiative
corrections to the neutrino mass operators ($O_M^{5D}$ and $O_M^{7D}$)
generated by the magnetic moment operators $O_W^-$ and $O_B$.  This allows us to determine constraints on the size
of the magnetic moment in terms of the neutrino mass, using
Eqs.(\ref{munu}) and (\ref{mnu}).  Our results are summarized in
Table~\ref{summary} below, where the quantity $R_{\alpha\beta}$ is defined as 
\begin{equation}
R_{\alpha\beta} = \left| \frac{m_\tau^2}{m_\alpha^2 - m_\beta^2} \right|,
\end{equation}
with $m_\alpha$ being the masses of charged lepton masses. Numerically, one has $R_{\tau e}
\simeq R_{\tau \mu} \simeq 1$ and $R_{\mu e} \simeq 283$.

\begin{table}[htbp]
   \centering
   \begin{tabular}{c|c|c}
\hline\hline
i) 1-loop, 7D & 
$\mu^W_{\alpha\beta}$ & $ \leq 1 \times 10^{-10}\mu_B
\left(\frac{ \left[m_\nu\right]_{\alpha\beta}}{1~{\rm eV}}\right)
\ln^{-1}\frac{\Lambda^2}{M_W^2} R_{\alpha\beta}$ \\
ii) 2-loop, 5D & $\mu^W_{\alpha\beta}$ & $ \leq 1 \times 10^{-9}\mu_B
\left(\frac{ \left[m_\nu\right]_{\alpha\beta}}{1~{\rm eV}}\right)
\left(\frac{1~{\rm TeV}}{\Lambda}\right)^2 
R_{\alpha\beta}$ \\  
\hline
iii) 2-loop, 7D & $\mu^B_{\alpha\beta}$ & $ \leq 1 \times 10^{-7}\mu_B
\left(\frac{ \left[m_\nu\right]_{\alpha\beta}}{1~{\rm eV}}\right)
\ln^{-1}\frac{\Lambda^2}{M_W^2}
R_{\alpha\beta}$ \\
iv) 2-loop, 5D &
$\mu^B_{\alpha\beta}$  &  $\leq 4 \times 10^{-9} \mu_B
\left(\frac{ \left[m_\nu\right]_{\alpha\beta}}{1~{\rm eV}}\right)
\left(\frac{1~{\rm TeV}}{\Lambda}\right)^2 
R_{\alpha\beta}$ \\
\hline\hline
   \end{tabular}
   \caption{Summary of constraints on the magnitude of the magnetic
moment of Majorana neutrinos.  The upper two lines correspond to a magnetic moment generated
by the $O_W$ operator, while the lower two lines correspond to the
$O_B^-$ operator.}
   \label{summary}
\end{table}


\section{SU(2) Gauge Boson}
\label{su2}

\begin{figure}[ht]
\includegraphics[width = 3.375in]{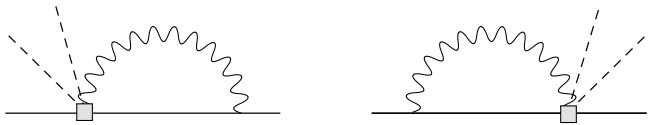}
\caption{Contribution of $O_W^+$ to the 5D neutrino mass operator.
The shaded box indicates the $O_W^+$ operator while solid, dashed, and
wavy lines correspond to leptons, Higgs, and gauge bosons, respectively.}
\label{fig:5D}
\end{figure}

We first consider contributions  from  $O_W^\pm$  to the mass operators.
Contributions to $O_M^{5D}$ occur through matching of the effective theory onto the full theory
at the scale $\Lambda$, illustrated at one-loop order with the diagrams of Fig.~\ref{fig:5D}.
If the operator $O_W^-$ is inserted at the
vertex, these two diagrams sum to zero.  This is no surprise, as these
two diagrams contain no non-trivial flavor dependence.  We
therefore cannot obtain a contribution to the flavor symmetric mass
operator by inserting the flavor antisymmetric operator $O_W^-$.

However, one-loop matching does yield a contribution to  $O_M^{5D}$ associated with the flavor symmetric operator  $O_W^+$ of order 
%
\begin{equation}
\label{C+5D}
C_M^{5D}(\Lambda) \simeq 
\frac{\alpha}{4 \pi \sin^2 \theta_W }
C_W^+(\Lambda)\ \ \ ,
\end{equation}
where we have estimated the one-loop matching contributions using
Fig.~\ref{fig:5D} and na\"ive dimensional analysis (NDA). We note that
this contribution arises from loop momenta of order $\Lambda$, and
that the precise numerical coefficient cannot be obtained without
knowing the structure of the theory above this scale. We therefore
follow the arguments of Ref.~\cite{Burgess:1992gx} and employ NDA to
estimate these high momentum contributions.

We see that the one-loop contribution to the 5D mass term provides a strong
constraint on $C_W^+$ but no constraint on the parameter $C_W^-$. 
In general, $C_W^\pm$
are unrelated parameters in the theory.  If the new physics were to
have no specific flavor symmetry/antisymmetry it might be natural for
$C_W^\pm$ to be of similar magnitude.  Alternatively, given the strong
constraint on $C_W^+$ arising from Eq.(\ref{C+5D}), a sizable magnetic
moment requires $|C_W^-| \gg |C_W^+|$. We identify two scenarios, to
be discussed below:
\begin{itemize}
\item
$|C_W^-| \sim |C_W^+|$
\item
Arbitrary $|C_W^-|$
\end{itemize}

\subsection{Special case: $C_W^- \sim C_W^+$}

We have seen that the flavor antisymmetric operator $O_W^-$ does not
contribute to the 5D neutrino mass term at 1-loop order; thus a direct
constraint on the magnetic moment is not obtained from the diagrams in
Fig.~\ref{fig:5D}.  However, suppose we had a theory in which the
coefficients of $O_W^+$ and $O_W^-$ were of similar magnitude, $C_W^+
\sim C_W^-$.  Then, using Eqs.(\ref{munu}), (\ref{mnu}) and
(\ref{C+5D}) we have
\begin{eqnarray}
m_\nu &\sim& \frac{\alpha}{8\pi \sin^2\theta_W} \frac{\Lambda^2}{m_e}
\frac{\mu_\nu}{\mu_B},
\nonumber \\
&\sim& \frac{\mu_\nu}{0.4 \times 10^{-15} \mu_B}
[\Lambda({\rm TeV})]^2 \,\,\,\,{\rm eV},
\label{special}
\end{eqnarray}
and thus obtain a stringent $\mu_\nu$ bound similar to that for Dirac
neutrinos.

We emphasize that Eq.(\ref{special}) is not a model-independent
constraint, as in general $O_W^+$ and $O_W^-$ are unrelated.  While it
might seem natural for the the new physics to generate coefficients of
similar size for both operators, we could obtain finite $C_W^-$ and
vanishing $C_W^+$ (at tree-level) by imposing an appropriate flavor
symmetry.

\subsection{General case: Arbitrary $C_W^-$.}
We now consider the more general case where $C_W^+$ and $C_W^-$ are
unrelated, and directly derive constraints on the the coefficient of
the flavor antisymmetric operator, $C_W^-$.

\subsubsection{7D mass term --- $O_W$}

\begin{figure}[ht]
\includegraphics[width = 2in]{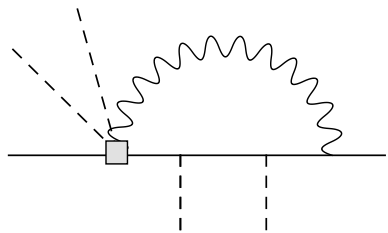}
\caption{Contribution of $O_W^-$ to the 7D neutrino mass operator.}
\label{fig:7D}
\end{figure}

As the operator $O_W^-$ is flavor antisymmetric, it must be
multiplied by another flavor antisymmetric contribution in order to
produce a flavor symmetric mass term.  This can be accomplished
through insertion of Yukawa couplings in the diagram shown in
Fig.~\ref{fig:7D}~\cite{Davidson}.  This diagram provides a
logarithmically divergent contribution to the 7D mass term, given by
\begin{equation}
\label{eq:owminusone}
\left[ C_M^{7D}(M_W)  \right]_{\alpha\beta}
\simeq \frac{3 g^2}{16 \pi^2}
\frac{m_\alpha^2 - m_\beta^2}{v^2} 
\ln \frac{\Lambda^2}{M_W^2}
\left[ C_W^-(\Lambda) \right]_{\alpha\beta}
\end{equation}
where $m_\alpha$ are the charged lepton masses, and the exact
coefficient has been computed using dimensional regularization and renormalized with modified minimal subtraction. We note that Eq.~(\ref{eq:owminusone}) gives the $O_W^-$ contribution to the neutrino mass  from all scales between $\Lambda$ and the scale of electroweak symmetry breaking to leading log order. Moreover, while all contributions of the type $(g^2 \ln \Lambda/M_W)^n$ could be resummed using the renormalization group, a study of analogous operator mixing for Dirac neutrinos\cite{dirac} suggests that it is sufficient to retain only the leading log contribution. Using this result -- as well as 
Eqs. (\ref{munu}) and (\ref{mnu}) to relate $C_W^-$ and $C_M^{7D}$
to $\mu_\mu$ and $m_\nu$ respectively --  leads to bound (i) in
Table~\ref{summary}.

Note that this provides a weaker constraint than that in
Eq.(\ref{special}), as it is suppressed by the charged lepton masses,
and also only logarithmically dependent on the scale of new physics
$\Lambda$.

\subsubsection{5D mass term --- $O_W$}

The neutrino magnetic moment operator $O_W^-$ will also
contribute to the 5D mass operator {\em via} two-loop matching of the effective theory onto the full theory at $\mu\sim\Lambda$.

\begin{figure}[h]
\includegraphics[width = 2in]{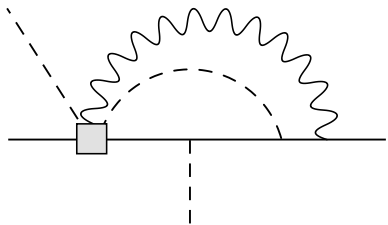}
\caption{Representative contribution of $O_W^-$ to the 5D neutrino
mass operator.}
\label{fig:2loop}
\end{figure}

An illustrative contribution is shown in Fig.~\ref{fig:2loop}.  As with the diagrams in
Fig.~\ref{fig:7D}, we require two Yukawa insertions in order to obtain
a flavor symmetric result.  This diagram contributes to
the 5D mass operator, and we again provide an NDA estimate:
\begin{equation}
\left[C_M^{5D}(\Lambda)  \right]_{\alpha\beta}
\simeq 
\frac{g^2}{(16 \pi^2)^2}
\frac{m_\alpha^2 - m_\beta^2}{v^2} 
\left[  C_W^- (\Lambda)\right]_{\alpha\beta}.
\label{5d1}
\end{equation}
Again, using Eqs.(\ref{munu}) and (\ref{mnu}), this leads to bound
(ii) in Table~\ref{summary}. In doing so, we have neglected the running of the operator coefficients from the scale $\Lambda$ to $M_W$ since the effects are higher order in the gauge couplings and have a negligible numerical impact on our analysis.

Compared to 1-loop (7D) case of Eq.~(\ref{eq:owminusone}), the 2-loop (5D) matching leads to a
mass contribution that is suppressed by a factor of $1/16\pi^2$ arising from the
additional loop, but enhanced by a factor of $\Lambda^2/v^2$ arising
from the lower operator dimension.  Thus, as we increase the new physics
scale, $\Lambda$, this two-loop constraint rapidly becomes more
restrictive and nominally provides a stronger constraint than the 1-loop result
once $\Lambda \agt 4 \pi v \sim 4$ TeV. Inclusion of the logarithmic $\Lambda$-dependence of one-loop mixing implies that the \lq\lq crossover " scale between the two effects is closer to $\sim 10$ TeV.


\section{Hypercharge Gauge Boson}
\label{u1b}

Unlike the case of the $SU(2)_L$ gauge boson, where a flavor symmetric
operator $O_W^+$ exists, the operator $O_B$ is purely flavor
antisymmetric.  Therefore, it cannot contribute to the $O_M^{5D}$ mass
term at one loop.  As was noticed in \cite{Davidson}, the one-loop
contribution of $O_B$ to the $O_M^{7D}$ mass term also vanishes.

\subsubsection{7D mass term --- $O_B$}

\begin{figure}[ht]
\includegraphics[width = 2in]{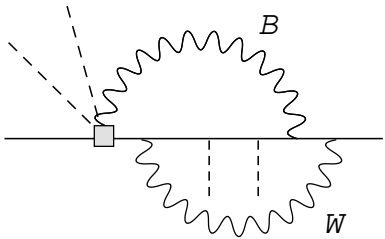}
\caption{Representative contribution of $O_B$ to the 7D neutrino
mass operator at two loop order.}
\label{fig:B2M_2loop}
\end{figure}

If we insert $O_B$ in the diagram in Fig. \ref{fig:7D}, the
contribution vanishes, due to the $SU(2)$ structure of the graph. 
Therefore, to obtain a non-zero contribution to $O_M^{7D}$ from
$O_B$ we require the presence of some non-trivial $SU(2)$ structure.  This can arise,  for instance, from a virtual $W$ boson loop as in
Fig. \ref{fig:B2M_2loop}~\cite{Davidson}.

This mechanism gives the leading contribution of the operator $O_B$ to
the 7D mass term.   The $O_B$ and $O_W$ contributions to the 7D mass term 
are thus related by
\begin{eqnarray}
\frac{(\delta m_\nu)^B}{(\delta m_\nu)^W}
\,\approx\,\frac{\alpha}{4\pi} \frac{1}{\cos^2\theta_W},
\end{eqnarray}
where $\theta_W$ is the weak mixing angle and where the factor on the RHS is due to the additional $SU(2)_L$ boson
loop.  This additional loop suppression for the $O_B$ contribution
results in a significantly weaker neutrino magnetic moment constraint
than that obtained above $O_W^-$.  The corresponding limit is shown as
bound (iii) in Table~\ref{summary}.

\subsubsection{5D mass term --- $O_B$}

However, the leading contribution of $O_B$ to the 5D mass term arises
from the same 2-loop matching considerations (Fig.~\ref{fig:2loop})
that we discussed in connection with the $O_W^-$ operator.  Therefore,
the contribution to the 5D mass term is the same as that for $O_W$,
except for a factor of $(g'/g)^2 = \tan^2 \theta_W$.  We thus obtain
\begin{equation}
\left[ C_M^{5D}(\Lambda) \right]_{\alpha\beta}
\simeq 
\frac{g'^2}{(16 \pi^2)^2}
\frac{m_\alpha^2 - m_\beta^2}{v^2} 
\left[ C_B(\Lambda)  \right]_{\alpha\beta}\ \ \ ,
\label{5db}
\end{equation}
corresponding to bound (iv) in Table.~\ref{summary}.  Importantly, this
is the strongest constraint on the $O_B$ contribution to the neutrino
magnetic moment for any value of $\Lambda$.


\section{Flavor Structure}
\label{flavor} 

All the magnetic moment bounds presented in Table~\ref{summary}
contain the factor $R_{\alpha\beta} = m_\tau^2/(m_\alpha^2 -
m_\beta^2)$.  This factor arises from the Yukawa insertions that are
necessary to obtain flavor symmetric mass terms from the flavor
antisymmetric magnetic moment operators.  The strongest constraints
are those for the $\mu_{\tau e}$ and $\mu_{\tau\mu}$ transition moments,
as $R_{\tau e} \simeq R_{\tau \mu} \simeq 1$.  However $R_{\mu e} =
283$, so the constraint on the $e-\mu$ transition moment is much
weaker.

In the foregoing analysis, it was convenient to work in the flavor
basis (where the charged lepton masses are diagonal).  However, given
$\mu_{\alpha\beta}$, we can use neutrino mixing matrix to determine
$\mu_{ij}$, where $i,j$ denote neutrino mass eigenstates.  Rotating
the neutrino magnetic moments from the flavor basis to the mass basis
gives
\begin{eqnarray}
\mu_{ij} &=& \sum_{\alpha\beta} \mu_{\alpha\beta} U^*_{\alpha i} U_{\beta j}
\\
& = & \mu_{\tau e} 
\left( U^*_{\tau i} U_{e j} - U^*_{e i} U_{\tau j} \right) 
+ \mu_{\tau \mu} ( ... ) +  \mu_{\mu e} ( ...).
\nonumber
\end{eqnarray}
Since most elements of $U_{i j}$ are large, it would be natural to
expect that all elements of $\mu_{ij}$ would be of similar size.  In
particular, all elements of $\mu_{ij}$ will receive a contribution
from $\mu_{\mu e}$ (the most weakly constrained element in the flavor
basis.)  Therefore the $\mu_{\mu e}$ constraint will translate into
similar limits for all elements of $\mu_{ij}$.


\section{Discussion and Conclusions}

We have discussed radiative corrections to the neutrino mass arising
from a neutrino magnetic moment coupling.  Expressing the magnetic
moment in terms of effective operators in a model independent fashion
required constructing operators containing the $SU(2)_L$ and hypercharge
gauge bosons, $O_W$ and $O_B$ respectively, rather than working
directly with the electromagnetic gauge boson.  We then calculated
$\mu_\nu$ naturalness bounds arising from the leading order contributions to both
the 5D and 7D Majorana mass terms.  These bounds are
summarized in Table~\ref{summary}.

At the TeV scale, the strongest bound comes from a 1-loop contribution
to the 7D mass term as previously discussed in~\cite{Davidson} [limit
(i) in Table~\ref{summary}].  However, this bound applies only to a
$\mu_\nu^W$ contribution to the magnetic moment, and not to a
$\mu_\nu^B$ contribution.  Even for the $\mu_\nu^W$ contribution, the
bound on the $e-\mu$ transition moment is weaker than current
experimental limits.

However, we have shown that 2-loop contributions to the 5D mass term
always provide the strongest bound on $\mu_\nu^B$ [limit (iv) in
Table~\ref{summary}] and also provides the strongest bound on
$\mu_\nu^W$ provided $\Lambda \agt 10$ TeV [limit (ii) in
Table~\ref{summary}].  We have estimated this contribution -- associated with matching the effective theory onto the (unspecified) full theory at high scales -- using NDA.
%
%
The resulting expressions for $\mu_\nu^B$ and $\mu_\nu^W$
differ only by a factor of $\tan^2\theta_W$.  Taking the $\mu_\nu^B$
limit (the weaker of the two) we have
\begin{equation}
\mu_{\alpha\beta}\,\leq\,4 \times 10^{-9}\mu_B 
\left(\frac{\left[m_\nu\right]_{\alpha\beta}}{1~{\rm eV}}\right)
\left(\frac{1~{\rm TeV}}{\Lambda}\right)^2 
\left| \frac{m_\tau^2}{m_\alpha^2 - m_\beta^2} \right|.
\label{generallimit}
\end{equation}
For any scale $\Lambda$, Eq.(\ref{generallimit}) is the most
general naturalness bound on the size of the Majorana neutrino magnetic moment.
It can only be evaded in the presence of fine tuning of the couplings or model-dependent suppression of the matching conditions at the scale $\Lambda$.

Turning now to the current experimental situation, the best laboratory
limit obtained from scattering of low energy reactor neutrinos is
$\text{``}\mu_{e}\text{''} < 0.7 \times 10^{-10}
\mu_B$~\cite{reactor}. Note that this limit applies to both $\mu_{\tau
e}$ and $\mu_{\mu e}$, as the flavor of the scattered neutrino is not
detected in the experiment. Taking the neutrino mass to be $m_\nu \alt
0.3$ eV (as implied by cosmological observations, e.g.\cite{WMAP3yr}),
Eq.(\ref{generallimit}) gives
\begin{eqnarray}
\mu_{\tau\mu},\mu_{\tau e} & \alt & 10^{-9} 
\left[ \Lambda(\text{TeV}) \right]^{-2}
\nonumber \\
\mu_{\mu e}  & \alt & 3 \times 10^{-7} 
\left[ \Lambda(\text{TeV}) \right]^{-2}.
\end{eqnarray}
We thus conclude that if $\mu_{\mu e}$ is dominant over the other
flavor elements, an experimental discovery near the present limits
(e.g., at $\mu \sim 10^{-11}\mu_B$) would imply that $\Lambda \alt 100$
TeV.  On the other hand, any model that leads to all elements of $\mu_{\alpha\beta}$ having similar size would only be consistent with the $m_\nu$ naturalness bounds if $\Lambda \alt 10$ TeV.

Our conclusions can be summarized according to the scale of the new
physics, $\Lambda$:
{\flushleft
I) $\Lambda \alt 10$ TeV 
\begin{itemize}
\item 
No conflict with experimental limits.
\item
Both $O_W$ and $O_B$ contributions to $\mu_\nu$ are possible, though
$O_W$ contributions are more tightly constrained.
\end{itemize}
II) $\Lambda \agt 10$ TeV
\begin{itemize}
\item 
$\mu_{\tau\mu}$, $\mu_{\tau e}$ bounds stronger than experimental limits
\item 
$\mu_{\mu e}$ bound weaker than experimental limits 
\item
Same limit irrespective of whether $\mu_\nu$ generated by $O_W$ and $O_B$
\end{itemize}
III) $\Lambda \agt 100$ TeV
\begin{itemize}
\item
The $\mu_{\alpha\beta}$ bound becomes stronger than current experimental constraints, for all $\alpha,\beta$.
\end{itemize}
}

The limit on the the magnetic moment of a Dirac neutrino is
considerably more stringent than for Majorana neutrino.  This is due
to the different flavor symmetries involved. In the Dirac case, no
insertion of Yukawa couplings is needed to convert a flavor
antisymmetric operator into a flavor symmetric operator, and the
stringent limit $\mu \leq 10^{-15} \mu_B$ holds (in the absence of
strong cancellations).  A significant implication is that if a
magnetic moment $\mu \agt 10^{-15} \mu_B$ were measured, it would
indicate that neutrinos are Majorana fermions, rather than
Dirac. Moreover, the scale of lepton number violation would be well
below the conventional see-saw scale.

\acknowledgments 

We thank Vincenzo Cirigliano, Concha Gonzalez-Garcia, and Mark Wise
for illuminating conversations.  This work was supported in part under
U.S. DOE contracts DE-FG02-05ER41361 and DE-FG03-92ER40701 and NSF
award PHY-0555674.  N.F.B was supported by a Sherman Fairchild
fellowship at Caltech.

\begin{appendix}
\section{Full 7 Dimensional Operator Basis for Majorana Neutrinos}

Here, we analyze the full set of gauge invariant operators of dimensions five through seven containing left-handed lepton doublet fields, one Higgs doublet, and $SU(2)_L$ and $U(1)_Y$ gauge bosons. It is convenient to classify the operators according to their dimension and field content. 

{\flushleft \it Mass operator (two fermion and four Higgs fields):}
\begin{equation}
O_{M}^{7D} = \left(\bar{L^c}\epsilon H\right)\left(H^T\epsilon
L\right) \left( H^\dagger H \right)
\end{equation}

{\flushleft \it Magnetic moment operators (two fermions, two Higgs and one gauge boson field):}
\begin{eqnarray}
O_B&=&\left(\bar{L^c}\epsilon H\right)\sigma_{\mu\nu}
\left(H^T\epsilon L\right)B^{\mu\nu}
\nonumber\\
O_W^a&=&\left(\bar{L^c}\epsilon H\right)\sigma_{\mu\nu}
\left(H^T\epsilon \tau^aL\right)W^{\mu\nu}_a
\nonumber\\
O_W^b&=&\left(\bar{L^c}\epsilon \tau^aH\right)\sigma_{\mu\nu}
\left(H^T\epsilon L\right)W^{\mu\nu}_a
\nonumber\\
O_W^c&=&\left(\bar{L^c}\epsilon \sigma_{\mu\nu}L\right)
\left(H\epsilon \tau^aH\right)W^{\mu\nu}_a
\end{eqnarray}
However, it can be shown that the operators involving $W^{\mu\nu}_a$
are related via
\begin{equation}
2O_W^a-O_W^c\;=\;-2O_W^b+O_W^c\;=\;O_W^a-O_W^b.
\end{equation}
Furthermore, we have 
\begin{equation}
\left[O_W^b\right]_{\alpha\beta}\,=\, \left[O_W^a\right]_{\beta\alpha}.
\end{equation}
Therefore, there is only one independent operator involving
$W^{\mu\nu}_a$, which we may choose to be $O_W^a \equiv O_W$.
We note that the operator defined in Ref.~\cite{Davidson},
\begin{equation}
O_W^{\rm Davidson} = \left(\bar{L^c}\epsilon\tau^a\sigma_{\mu\nu}L\right)
\left(H\epsilon\tau^bH\right)W^{\mu\nu}_d\varepsilon_{abd},
\end{equation}
corresponds to the flavor antisymmetric component of  $O_W$.  Explicitly, 
$O_W^{\rm Davidson} = O_W^a - O_W^b \equiv 2 O_W^-$.

{\flushleft\it Two fermion fields, two Higgs fields, and two derivatives:}
\begin{eqnarray}
O_3^a&=&\left(\bar{L^c}\epsilon H\right)\left(D_\mu H^T\epsilon D^\mu L\right)
\nonumber\\
O_3^b&=&\left(D_\mu \bar{L^c}\epsilon D^\mu H\right)\left(H^T\epsilon L\right)
\nonumber\\
O_3^c&=&\left(D_\mu \bar{L^c}\epsilon H\right)\left(H^T\epsilon D^\mu L\right)
\nonumber\\
O_3^d&=&\left(\bar{L^c}\epsilon D^\mu H\right)\left(D_\mu H^T\epsilon L\right)
\end{eqnarray}
It is also possible to establish relationships between some of the 
$O_3^i$ operators. Transposition yields:
\begin{eqnarray}
\left(O_3^a\right)^T
&=&-\left(D_\mu L^T\epsilon^T D^\mu H\right)\left(H^T\epsilon^T C^TL\right)
\nonumber\\
&=&O_3^b,
\end{eqnarray}
where the minus sign arises from the interchange of two fermion fields, 
and we use $\epsilon^T=-\epsilon$, $C^T=-C$. Since the transpose of a number 
is the same number again, we conclude that there is only one independent 
operator out of these two, which we may choose to be $O_3^a\, \equiv \,O_3^1$.

The same trick applied to the remaining two operators does not lead to
any new relations. However, another relation can be obtained via
integration by parts.  We consider
\begin{equation}
0 = D_\mu\left(D_\mu \bar{L^c}\epsilon H H^T\epsilon L\right)
\,\equiv\,\{11\}+\{12\}+\{13\}+\{14\},
\end{equation}
where the numbers in square brackets denote the field on which the first and 
the second derivative act, respectively. 
The squared derivative acting on Higgs is zero by the equation of motion, 
and that acting on fermion fields leads to operators of the form $O_{W,B}$.
Repeating this trick for all possible combinations, we obtain a system of 4 
equations:
\begin{eqnarray}
\{11\}+\{12\}+\{13\}+\{14\}&=&0\\
\{12\}+\{23\}+\{24\}&=&0\nonumber\\
\{13\}+\{23\}+\{34\}&=&0\nonumber\\
\{14\}+\{24\}+\{34\}+\{44\}&=&0\nonumber\\
\frac{1}{2}(\{11\}+\{44\})+\{12\}+\{13\}&&\nonumber\\
+\{14\}+\{23\}+\{24\}+\{34\}&=&0,\nonumber
\end{eqnarray}
where the last equation is the sum of the first four. Subtracting from this 
5th equation the 1st and the 4th, we obtain:
\begin{eqnarray}
\{14\}\;=\;\{23\}-\frac{1}{2}(\{11\}+\{44\}),
\end{eqnarray}
which implies that the operator $O_3^c$ can be expressed in terms of the 
operators $O_3^d$ and $O_{W,B}$. 
We are thus left with just two independent operators with two derivatives:
\begin{eqnarray}
O_3^1&=&\left(\bar{L^c}\epsilon H\right)\left(D_\mu H^T\epsilon D^\mu L\right)
\nonumber\\
O_3^2&=&\left(\bar{L^c}\epsilon D_\mu H\right)\left(D^\mu H^T\epsilon L\right).
\end{eqnarray}

{\flushleft\it Two fermion fields, three Higgs fields, and one derivative:}
\begin{eqnarray}
O_4^a&=&\bar{e^c}\gamma_\mu\left(H^T\epsilon \tau^aL\right)
\left(D^\mu H^T\epsilon \tau_aH\right)
\nonumber\\
O_4^b&=&\bar{e^c}\gamma_\mu\left(D^\mu H^T\epsilon \tau^aL\right)
\left(H^T\epsilon \tau_aH\right)
\end{eqnarray}

After fierzing, both operators reduce to just one independent operator:
\begin{eqnarray}
O_4&=&\bar{e^c}\gamma_\mu\left(H^T\epsilon L\right)
\left(D^\mu H^T\epsilon H\right)
\end{eqnarray}

{\flushleft\it Four fermion fields and one Higgs:}
\begin{eqnarray}
O_5^S&=&\left(\bar{L^c}\epsilon L\right)
\left(L^T_ie^c\right)\epsilon_{ij}H_j
\nonumber\\
O_5^T&=&\left(\bar{L^c}\epsilon \sigma_{\mu\nu}L\right)
\left(L^T_i\sigma^{\mu\nu}e^c\right)\epsilon_{ij}H_j
\end{eqnarray}

Thus $\left\{O_{M}^{7D} , O_B, O_W, O_3^1, O_3^2, O_4, O_5^S, O_5^T
\right\}$ form the complete set of 7D operators (where flavor labels have been suppressed).

\end{appendix}


\begin{thebibliography}{99}

\bibitem{Marciano:1977wx}
W.~J.~Marciano and A.~I.~Sanda,
Phys.\ Lett.\ B {\bf 67}, 303 (1977);
%
B.~W.~Lee and R.~E.~Shrock,
Phys.\ Rev.\ D {\bf 16}, 1444 (1977);
%
K.~Fujikawa and R.~Shrock,
Phys.\ Rev.\ Lett.\  {\bf 45}, 963 (1980).

\bibitem{Beacom} 
 J.~F.~Beacom and P.~Vogel,
Phys.\ Rev.\ Lett.\  {\bf 83}, 5222 (1999);
D.~W.~Liu {\it et al.},
Phys.\ Rev.\ Lett.\  {\bf 93}, 021802 (2004).


\bibitem{reactor} 
H.~T.~Wong  [TEXONO Collaboration],
  hep-ex/0605006.
B.~Xin {\it et al.}  [TEXONO Collaboration],
  Phys.\ Rev.\ D {\bf 72}, 012006 (2005);
Z.~Daraktchieva {\it et al.}  [MUNU Collaboration],
  Phys.\ Lett.\ B {\bf 615}, 153 (2005).


\bibitem{Raffelt} G.G. Raffelt, Phys. Rep. {\bf 320}, 319 (1999).


\bibitem{Fukugita}
  M.~Fukugita and T.~Yanagida, 
{\it Physics of neutrinos and applications to astrophysics},
Chapter 10, Springer, Berlin, (2003), and references therein.

\bibitem{boris}
  B.~Kayser, F.~Gibrat-Debu and F.~Perrier,
  World Sci.\ Lect.\ Notes Phys.\  {\bf 25}, 1 (1989).

\bibitem{Wong}
  H.~T.~Wong and H.~B.~Li,
  Mod.\ Phys.\ Lett.\ A {\bf 20}, 1103 (2005).

\bibitem{McLaughlin}
G.~C.~McLaughlin and J.~N.~Ng,
  Phys.\ Lett.\ B {\bf 470}, 157 (1999);
  R.~N.~Mohapatra, S.~P.~Ng and H.~b.~Yu,
  Phys.\ Rev.\ D {\bf 70}, 057301 (2004).

\bibitem{Balantekin:2006sw}
  A.~B.~Balantekin,
  hep-ph/0601113;
  A.~de Gouvea and J.~Jenkins,
  hep-ph/0603036;
  A.~Friedland,
  hep-ph/0505165.

\bibitem{Voloshin}
  M.~B.~Voloshin,
  Sov.\ J.\ Nucl.\ Phys.\  {\bf 48}, 512 (1988).
For a specific implementation, see 
  R.~Barbieri and R.~N.~Mohapatra,
  Phys.\ Lett.\ B {\bf 218}, 225 (1989).


\bibitem{Georgi}
  H.~Georgi and L.~Randall,
  Phys.\ Lett.\ B {\bf 244}, 196 (1990).

\bibitem{Grimus}
  W.~Grimus and H.~Neufeld,
  Nucl.\ Phys.\ B {\bf 351}, 115 (1991).

\bibitem{mohapatra1}
  K.~S.~Babu and R.~N.~Mohapatra,
  Phys.\ Rev.\ Lett.\  {\bf 64}, 1705 (1990).

\bibitem{Barr}
  S.~M.~Barr, E.~M.~Freire and A.~Zee,
  Phys.\ Rev.\ Lett.\  {\bf 65}, 2626 (1990).

\bibitem{dirac}
  N.~F.~Bell, V.~Cirigliano, M.~J.~Ramsey-Musolf, P.~Vogel and M.~B.~Wise,
  Phys.\ Rev.\ Lett.\  {\bf 95}, 151802 (2005).

\bibitem{Davidson}
  S.~Davidson, M.~Gorbahn and A.~Santamaria,
  Phys.\ Lett.\ B {\bf 626}, 151 (2005).

\bibitem{Cirigliano}
  V.~Cirigliano, A.~Kurylov, M.~J.~Ramsey-Musolf and P.~Vogel,
  Phys.\ Rev.\ Lett.\  {\bf 93}, 231802 (2004).


\bibitem{mohapatra2}
  K.~S.~Babu and R.~N.~Mohapatra,
  Phys.\ Rev.\ Lett.\  {\bf 63}, 228 (1989).

\bibitem{Burgess:1992gx}
  C.~P.~Burgess and D.~London,  
  Phys.\ Rev.\ D {\bf 48}, 4337 (1993).

\bibitem{WMAP3yr}
  D.~N.~Spergel {\it et al.},
  astro-ph/0603449.





\end{thebibliography}
\end{document}